# *Plasmodium knowlesi* H strain pregnancy malaria immune responses in olive baboons (*Papio anubis*)

**Barasa Mustafa[1,2*], Mwangi Irungu Michael[1,2], Mutiso Muli Joshua[1,2], Kagasi Ambogo Esther[2], Ozwara Suba Hastings[2], Gicheru Muita Michael[1]**

[1]Zoological Sciences Department, School of Pure and Applied Sciences, Kenyatta University, Nairobi
Department of Tropical and Infectious Diseases, Institute of Primate Research, Nairobi, Kenya

**ABSTRACT**

Approximately 24 million pregnant women in Sub-Saharan Africa are at risk of suffering from pregnancy malaria complications. Mechanisms responsible for increased susceptibility to malaria in pregnant women are not fully understood. Baboons are susceptible to *Plasmodium knowlesi* and their reproductive physiology and host pathogen interactions are similar to those in humans, making them attractive for development as a model for studying mechanisms underlying pregnancy malaria. This study exploited the susceptibility of baboons to *Plasmodium knowlesi* infection to characterize cytokine and peripheral blood mononuclear cell recall proliferation responses underlying the pathogenesis of pregnancy malaria in baboons infected with *Plasmodium knowlesi*. The pregnancies of three time mated adult female baboons and their gestational levels were confirmed by ultrasonography. On the 150th day of gestation, the pregnant baboons together with four non pregnant controls were infected with *Plasmodium knowlesi* H strain parasites. Collection of peripheral sera, and mononuclear cells was then done on a weekly basis. Sera cytokine concentrations were measured by Enzyme Linked Immunosorbent Assay (ELISA) using respective enzyme conjugated antibodies. Peripheral blood mononuclear cell recall proliferation assays were also done on a weekly basis. Results indicate that pregnancy malaria in this model is associated with suppression of interferon gamma and interleukin 6 (IL-6) responses. Tumour necrosis factor alpha responses were upregulated while IL-4, IL-12 and recall proliferation responses were not different from controls. These data to a great extent are consistent with some findings from human studies, showing the feasibility of this model for studying mechanisms underlying pregnancy malaria.

**Keywords**: Baboon; pregnancy; malaria; cytokine; proliferation.

## INTRODUCTION

Several experiments have demonstrated that there is increased susceptibility to *P. falciparum* malaria infection in pregnant women (Menendez *et al.,* 2000). Malaria during pregnancy leads to many complications in women and their infants, threatening the lives of both the mother and the child. These complications include stillbirths, abortions, pre-term births, low birth weights, reduction in gestational age, anaemia and high fever (Menendez *et al.,* 2000; Steketee *et al.,* 2001). Similar pregnancy malaria complications were demonstrated in the first baboon (*Papio anubis*) model of placental malaria

**\***Corresponding author:
*Mustafa Barasa, MSc.*
*Zoological sciences Department,*
*School of Pure and Applied Sciences,*
*P. O. Box 43844, Kenyatta University, Nairobi, Kenya*
*Email: mustrech@yahoo.com*

(Barasa *et al.,* 2010). An estimated 24 million pregnant women in sub Saharan Africa are at risk of suffering from pregnancy malaria and prevalence may exceed 50% among primigravidae and secundigravidae in endemic areas (Phillips-Howard *et al.,* 1999). There is no vaccine and currently one of the best therapeutic measures involves the use of artemisinin based combinational therapy.

Even though pregnancy malaria is to an extent studied in pregnant women, the studies have shortcomings resulting from confounding inherent variables such as the mother's health status, inaccurate estimation of gestational age, inadequate tissue for analysis, patient compliance problems, socio-economic conditions, moral, ethical and financial limitations (Moore *et al.,* 1999; Steketee *et al.,* 1996). As a result, many questions are not satisfactorily addressed during human studies. To our knowledge, apart from TNF-α (Davison *et al.,* 2006), immunoglobulin G and immunoglobulin

M (Barasa *et al.,* 2010), no other immunological parameters had been characterized in the non human primate models of pregnancy malaria. This study aimed to characterize immunological mechanisms (cytokine and recall proliferation responses) underlying pregnancy malaria in the baboon-*P. knowlesi* nonhuman primate model.

## MATERIALS AND METHODS

### Experimental animals and study design

Three adult female baboons were maintained in the baboon colony facility at the Institute of Primate Research (IPR) in the company of an adult male baboon for mating to occur. Ultrasound tests were used to confirm pregnancy status and gestation periods of the baboons. The baboons were infected together with four non pregnant control baboons with *Plasmodium knowlesi* blood stage, overnight cultured parasites
(Barasa *et al.,* 2010) on the gestation day 150. Each baboon received an inoculum of 1.0x106 parasites/ml in incomplete RPMI 1640 (Sigma-Aldrich, USA). Following infection the baboons were bled on a weekly basis for extraction of sera and plasma for ELISA assays. Plasma samples were used for TNF-α ELISA assays. At 5% level of parasitaemia baboons were intravenously injected with chloroquin sulphate at a dosage of 5 mg /kg body weight daily for 3 days. Baboon groups used were as follows: (1) Pregnant infected; PAN 2724 (PAN means *Papio anubis*), PAN 2809, PAN 2859 and (2) Non pregnant infected; PAN 2870, PAN 3023, PAN 2911 and PAN 3035. For all invasive procedures, the baboons were anaesthetised with ketamine hydrochloride (10 mg/kg body weight).

### Determination of cytokine responses in sandwich ELISA

Cytokine ELISA assays were done in order to measure the levels of cytokine mediated immune responses induced in the baboons. Ninety six well flat bottomed ELISA microtiter plates (Sigma-Aldrich, USA) were coated with 5 μg/ml of cross reactive antihuman/baboon cytokine (IFN-γ, TNF-α, IL-4, IL-6,
IL-12) capture monoclonal antibodies (Becton Dickinson) delivered 50 μl/well. These were incubated overnight at 4ºC. Excess coating buffer was then flicked off and the wells blocked with 100 μl/well blocking buffer (3% BSA in PBS) followed by 1 hr incubation at 37°C. After washing the plates six times using ELISA washing buffer (0.05% tween in PBS), undiluted sera samples and recombinant cytokine standards were dispensed in duplicate, 50 μl/well and plates incubated for 2 hr at 37°C. Standards were serial diluted by transferring 50 μl from well to well with mixing beginning with a neat concentration of 500 pg/ ml.

The plates were then washed as before and detector mouse biotinylated antibaboon cytokine (IFN-γ, TNF-α IL-4, IL-6, IL-12) monoclonal antibodies (Becton Dickinson) added 50 μl/well at dilutions of 1:2000. This was followed by a one hr incubation period at 37°C then washing as before. Streptavidin Horse Radish Peroxidase (HRP) diluted 1:2000 was added 50 μl/well and incubated 1 hr at 37°C followed by washing six times. Colour development was achieved by adding 50 μl/well of Tetramethylbenzidene (TMB) substrate and optical densities were read using a Dynatech MRX ELISA reader at 630 nm filter setting after 15 min of incubation at 37°C (Barasa *et al.,* 2010; Gicheru *et al.,* 1995).

### Peripheral blood mononuclear cells recall proliferation assays

Proliferation assays were necessary for the determination of the infected baboons' cell mediated specific, acquired immune responses. Ninety six well round bottom microtitre plates were used for the assays. Peripheral Blood Mononuclear Cells (PBMC) were resuspended for delivery of 2x105cells/well dispensed in 100 μl of complete RPMI 1640 (RPMI with 10% FBS, 2 mM L-glutamine, 100μg/ml gentamycin, 0.05 mM 2-mercaptoethanol). The cells were stimulated in triplicate wells with 105 *Plasmodium knowlesi* crude parasite antigen per well or 10μg/ml final concentration of Concanavalin A (Con A; for positive control). Triplicate control background wells received 50μl of complete media. The plates were then covered and taped around to prevent rapid evaporation. Cultures were incubated at 37°C in a humidified incubator for 5 days for *Plasmodium knowlesi* antigen cultures and for 3 days for Con A cultures. Cells were pulsed with 0.5 μCi of [methyl-3] thymidine (20 μl delivery/well) over the last 18 hours then harvested on a fiber filter paper. Radionuclide incorporation was expressed in form of stimulation indices (Gicheru *et al.,* 1997; Gicheru *et al.,* 2001; Olobo *et al.,* 1995).

### Statistical analysis

Mean values of cytokine levels and recall proliferation stimulation indices (SI) of pregnant infected group of baboons were compared with the values of control group of baboons (non pregnant infected baboons) using non parametric Mann-Whitney U analysis. Probability values of *P* < 0.05 were considered significant.

## RESULTS

### Th1 cytokine responses

Non pregnant baboons had slightly higher baseline levels of IFN-γ levels than pregnant ones (55.54 pg/ml for pregnant and 62.30 pg/ml for non pregnant ones).

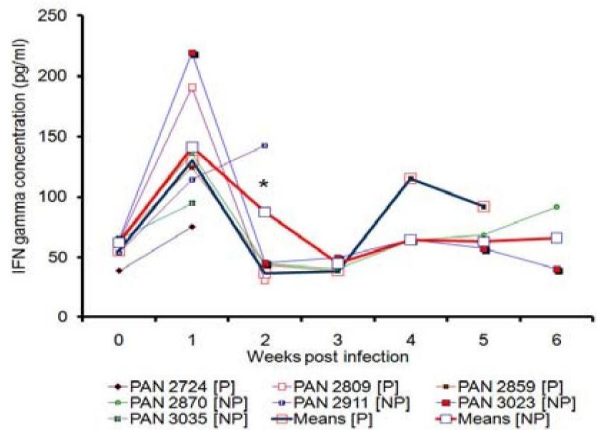

**Figure 1:** IFN-γ responses in *P. knowlesi* infected baboons. PAN: *Papio anubis*; P: Pregnant; NP: Non pregnant. *: Asterisk means significant difference between pregnant and non pregnant baboons in IFN-γ responses, P < 0.05; n: P = 3; NP = 4.

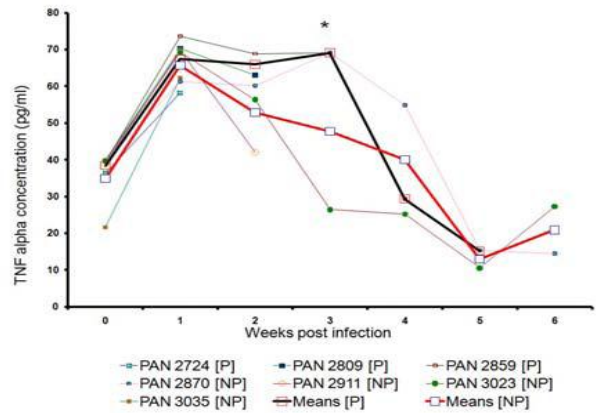

**Figure 2:** Tumour necrosis factor alpha responses in *P. knowlesi* infected baboons. PAN: *Papio anubis*; P: Pregnant; NP: Non pregnant. *: Asterisk means significant difference between pregnant and non pregnant baboons in TNF α responses, P < 0.05; n: P = 3; NP = 4.

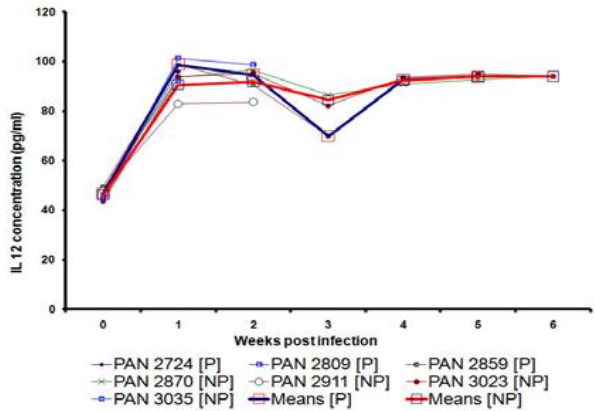

**Figure 3:** Interleukin 12 responses in *P. knowlesi* infected baboons. PAN: *Papio anubis*; P: Pregnant; NP: Non pregnant. n: P=3; NP = 4.

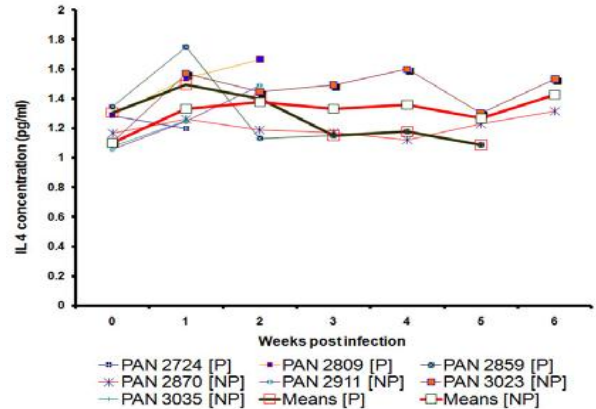

**Figure 4:** Interleukin 4 responses in *P. knowlesi* infected baboons. PAN: *Papio anubis*; P: Pregnant; NP: Non pregnant. n: P = 3; NP= 4.

Following infection both groups had sharp increases in IFN-γ concentration peaking at week one post infection. In pregnant baboons concentration of IFN-γ ranged from 36.81 pg/ml to 9.92 pg/ml while in non pregnant baboons levels ranged from 54.98 pg/ml to 141.15 pg/ml. Non pregnant baboons produced significantly higher levels of IFN-γ than pregnant baboons during the early phase of infection (Fig. 1; P < 0.05). Baseline production of TNF alpha was similar between the pregnant and non pregnant group of baboons (38.46 pg/ml in pregnant baboons and 34.92 pg/ml in non pregnant baboons). In pregnant baboons TNF alpha concentration ranged from 15.18 pg/ml to 69.10 pg/ml while in non pregnant baboons it ranged from 12.97 pg/ml to 55.62 pg/ml after infection. Pregnant baboons produced significantly higher levels of TNF alpha than non pregnant baboons (Fig. 2; P < 0.05). Mean baseline

IL-12 levels were similar between the pregnant and non pregnant baboons in the study (46.94 pg/ml in pregnant baboons and 46.16 pg/ml in non pregnant baboons). Following infection both groups experienced rapid increases in IL-12 concentrations (Fig. 3). In pregnant baboons IL-12 levels ranged from 46.94 pg/ml to 98.64 pg/ml while in non pregnant baboons levels ranged from 46.16 pg/ml to 94.00 pg/ml over the six weeks of experimentation. When IL-12 means were considered, the differences between pregnant and non pregnant baboons were not significant (Fig. 3; P > 0.05).

**Th2 cytokine responses**

Pregnant baboons had higher mean baseline interleukin 4 levels than their non pregnant counterparts (1.31 pg/ml compared to 1.10 pg/ml). After infection levels of IL-4 in pregnant baboons ranged from 1.09 pg/ml to 1.49 pg/ml while in non pregnant baboons levels ranged from 1.10 pg/ml to 1.42 g/ml. Comparison of the mean production of IL-4 between pregnant and non pregnant baboons showed that differences between the two groups were not significant (Fig. 4; P > 0.05). Pregnant and non pregnant baboons produced similar baseline levels of IL-6 (3.53 pg/ml and 3.56 pg/ml respectively). After infection, pregnant baboons produced IL-6 levels ranging from 2.98 pg/ml to 6.78 g/ml while in non pregnant baboons levels ranged from 3.46 pg/ml to 11.16 pg/ml. Non pregnant baboons produced significantly higher IL-6 levels than pregnant baboons through out the period of experimentation (Fig. 5; P < 0.05).

### Peripheral blood mononuclear cells recall proliferation responses

Proliferation assays were carried out beginning with baseline assays then every week post infection for up to six weeks. Baseline stimulation indices detected were comparable in the pregnant and non pregnant baboons. Proliferation in pregnant baboons ranged between 2.67 and 4.87 while in non pregnant baboons it ranged between 2.47 and 6.11. When mean proliferation were considered, the differences between pregnant and non pregnant baboons were not significant (Fig. 6; P > 0.05).

### DISCUSSION

The aim of this work was to determine the immunological mechanisms (IFN-γ, TNF-α IL-4, IL-12 levels and recall proliferation responses) underlying pregnancy malaria in the baboon - *P. knowlesi* model. Non pregnant baboons produced significantly higher IFN-γ levels early during the experimentation than pregnant baboons. This outcome may be attributed to the downregulation of IFN-γ mediated immunity in malaria during pregnancy in the baboon model. Malaria-infected individuals produce large amounts of proinflammatory cytokines, such as IFN-γ and TNF-α. This innate cytokine response is responsible for the high levels of fever that occur within a few days of the onset of blood stage infection in nonimmune individuals (Davison *et al.,* 2006). Tumour necrosis factor alpha levels in peripheral blood were significantly increased in pregnant baboons than in non pregnant ones early during the infection. The condition of pregnancy conferred an upregulation of TNF-α levels following infection with *P. knowlesi*. In a similar rhesus monkey model study TNF-α was upregulated in pregnant rhesus monkeys that had been infected with *P. Coatneyi* (Davison *et al.,* 2006). In a recent human study, IFN-γ was found to be important in the protection against placental malaria. *Plasmodium* antigen-stimulated placental intervillous blood mononuclear cells (IVBMC) from multigravids with less placental infection were found to secrete more interferon gamma than IVBMC from primigavids or secundigravids (Moore *et al.,* 1999). Concentrations of IL-12 detected in pregnant and non pregnant baboons were not significantly different.

Therefore the pregnancy status of baboons did not alter the level of IL-12 immune responses when pregnant baboons were infected with *P. knowlesi*.

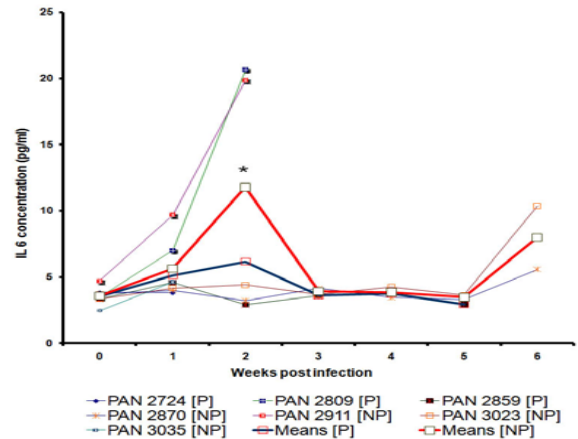

**Figure 5:** Interleukin 6 responses in *P. knowlesi* infected baboons. PAN: *Papio anubis*; P: Pregnant; NP: Non pregnant. *: Asterisk means significant difference in IL 6 responses between pregnant and non pregnant baboons, P < 0.05. n: P = 3; NP = 4.

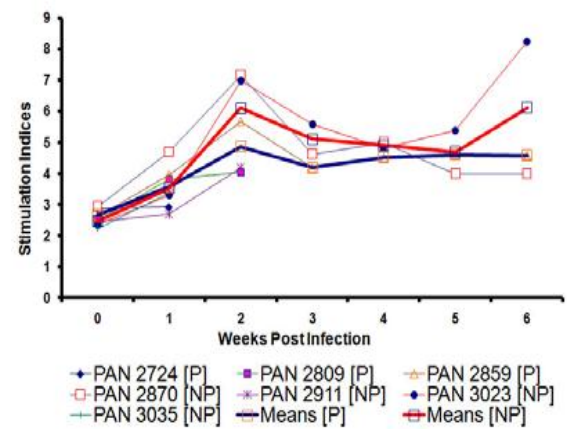

**Figure 6:** Mean weekly PBMC proliferation observed in *P. knowlesi* infected baboons. PAN: *Papio anubis*; P: Pregnant; NP: Non pregnant. n: P = 3; NP = 4.

In this study, pregnant baboons produced similar levels of IL-4 to non pregnant baboons following infection. Pregnancy (during the last trimester) is generally associated with a bias towards T helper 2 immunity in order to maintain the conceptus as an immunologically privileged site (Saito, 2000). In the current experiment, IL-4 immune responses were not affected by the pregnancy status of baboons infected with *P. knowlesi*. Non pregnant baboons produced significantly higher levels of IL-6 than pregnant ones throughout the period of experimentation. This result shows that IL-6 mediated immunity is downregulated in pregnant baboons during malaria infection.

Pregnant baboons had comparable stimulation indices (SI) to non pregnant baboons, showing that PBMC proliferation against the *P. knowlesi* crude antigen was not affected by gravidity. Previous studies have shown that peripheral immunity normally remains as competent in pregnant women as it is in non pregnant ones against infectious agents (Fievet *et al.,* 2001). In the mouse model for pregnancy malaria, pregnant *P. chabaudi* infected mice were found to have comparable splenocyte proliferation responses compared to non pregnant mice on day 6 post infection (Jayekumar and Moore, 2006). The findings from the current study, together with the first characterisation of placental malaria in olive baboons (*Papio anubis*) infected with *P. knowlesi* H strain parasites (Barasa *et al.,* 2010), could have far reaching implications since *P. knowlesi* has recently become a human parasite (Ng *et al.,* 2008). Data from the current study suggests that while IFN-γ and IL-6 immunity is downregulated TNF-α immunity is upregulated during pregnancy malaria in the baboon – *P. knowlesi* model. Proliferation, IL-12, and IL-4 responses remain largely unchanged during pregnancy malaria in this newly established non-human primate model of pregnancy malaria.

## ACKNOWLEDGEMENT

This study was funded by the research capability strengthening WHO grant (Grant Number: A 50075) for malaria research in Africa under the Multilateral Initiative on Malaria/Special Programme for Research and Training in Tropical Diseases (WHO-MIM/TDR). We are grateful to the entire Animal Resources Department at the Institute of Primate Research (IPR) for providing the baboons and other support during the study.